\documentclass[aps,prl,twocolumn,superscriptaddress]{revtex4}
\usepackage{ucs}
\usepackage[warn]{mathtext}
\usepackage[utf8x]{inputenc}
\usepackage[english]{babel}
\usepackage[xdvi]{graphicx}
\usepackage{amssymb,amsmath}
\begin{document}
\title{Magnetic properties of RCoO$_3$ cobaltites (R = La, Pr, Nd, Sm, Eu).\\
Effects of hydrostatic and chemical pressure.}

\author{A.S. Panfilov}
\email[]{panfilov@ilt.kharkov.ua}
\affiliation{B. Verkin Institute for Low Temperature Physics and Engineering,
National Academy of Sciences of Ukraine, 61103 Kharkov, Ukraine}

\author{A.A. Lyogenkaya}
\affiliation{B. Verkin Institute for Low Temperature Physics and Engineering,
National Academy of Sciences of Ukraine, 61103 Kharkov, Ukraine}

\author{G.E. Grechnev}
\email[]{grechnev@ilt.kharkov.ua}
\affiliation{B. Verkin Institute for Low Temperature Physics and Engineering,
National Academy of Sciences of Ukraine, 61103 Kharkov, Ukraine}

\author{V.A. Pashchenko}
\affiliation{B. Verkin Institute for Low Temperature Physics and Engineering,
National Academy of Sciences of Ukraine, 61103 Kharkov, Ukraine}

\author{I.P. Zhuravleva}
\affiliation{B. Verkin Institute for Low Temperature Physics and Engineering,
National Academy of Sciences of Ukraine, 61103 Kharkov, Ukraine}

\author{L.O. Vasylechko}
\affiliation{Lviv Polytechnic National University, 79013 Lviv, Ukraine}

\author{Vasyl Hreb}
\affiliation{Lviv Polytechnic National University, 79013 Lviv, Ukraine}

\author{V.A. Turchenko}
\affiliation{Joint Institute for Nuclear Research, 141980 Dubna, Russia}

\author{D. Novoselov}
\affiliation{M.N. Mikheev Institute of Metal Physics,\\
Ural Branch of the Russian Academy of Sciences, 620108 Yekaterinburg, Russia}
\affiliation{Ural Federal University, 620002 Yekaterinburg, Russia}

\begin{abstract}
We have investigated the temperature dependence of the magnetic susceptibility $\chi(T)$
of rare-earth cobaltites RCoO$_3$ (R= La, Pr, Nd, Sm, Eu) in the temperature range
$4.2-300$ K and also the influence of hydrostatic pressure up to 2 kbar on their
susceptibility at fixed temperatures $T=78 $ and 300 K.
The specific dependence $\chi(T)$ observed in LaCoO$_3$ and the anomalously large
pressure effect (d\,ln $\chi$/d$P\sim -100$ Mbar$^{-1}$ for $T = 78$ K) are
analysed in the framework of a two-level model with energy levels difference $\Delta$.
The ground state of the system is assumed to be nonmagnetic with the zero spin of Co$^{3+}$ ions,
and magnetism at a finite temperature is determined by the excited magnetic spin state.
The results of the analysis, supplemented by theoretical calculations of the
electronic structure of LaCoO$_3$, indicate a significant
increase in $\Delta$ with a decrease in the unit cell volume under the hydrostatic pressure.
In the series of RCoO$_3$ (R= Pr, Nd, Sm, Eu) compounds, the volume of crystal cell
decreases monotonically due to a decrease in the radius of R$^{3+}$ ions.
This leads to an increase in the relative energy $\Delta$ of the excited state
(the chemical pressure effect), which manifests itself in a decrease in the contribution
of cobalt ions to the magnetic susceptibility at a fixed temperature, and also
in a decrease in the hydrostatic pressure effect on the susceptibility of
RCoO$_3$ compounds, which we have observed at $T=300$~K.
\end{abstract}



\maketitle

\section{Introduction}
It is well known that Co$^{3+}$ ions with $3d^6$ electronic configuration in RCoO$_3$
cobaltites, which have a perovskite-like crystal structure, can exist in three
different spin states corresponding to the low (LS, $S=0$), intermediate (IS, $S=1$),
and high (HS, $S=2$) spin values.
The energies of these states are determined by the competition between the splitting
of ionic energy levels by the crystal field in $t_{2g}$ and $e_g$ states and
Hund's intra-atomic exchange interaction.
As a result, the states with different spin of Co$^{3+}$ ions have comparable energy values,
and their relative positions are sensitive to such factors as temperature,
pressure and magnetic field.
These factors can generate different spin crossovers and provide the observed peculiar
behaviour and variety of physical properties of the RCoO$_3$ systems
(see Refs. \cite{Raveau12,Takami14,Raveau15} and references therein).

The evolution of the spin states of Co $^{3+}$ with increasing temperature is most
clearly manifested in the magnetic properties of  LaCoO$_3$ compound, where La
does not have a magnetic moment and the contribution of cobalt ions to the
susceptibility $\chi$ appears to be predominant.
It is recognized that in the ground state the cobalt ions are in the low-spin LS state ($S$=0)
and at low temperatures LaCoO$_3$ is a nonmagnetic semiconductor.
As temperature rises, the states with a higher spin, IS and/or HS, begin to populate.
This leads to a rapid increase in the susceptibility and a pronounced maximum in
$\chi(T)$ at $T\sim 100$ K, followed by a decrease in susceptibility close to
the Curie-Weiss law \cite{Zobel02,Baier05,Yan04}.
In addition, another feature in $\chi(T)$ with the shape of a wide plateau is observed
near 500 K, which is associated with a transition to the metallic type of conductivity.

Despite the huge amount of experimental and theoretical works devoted to studies of
RCoO$_3$ cobaltites, a nature of the spin state of Co$^{3+}$ ions and mechanisms of its
manifestation in physical properties is the subject of continuous discussions
(see e.g. Ref. \cite{Rao04} and references therein).
One of the main questions concerns the scenario of transition between spin states of
cobalt with increasing temperature: LS $\to$ IS, LS $\to$ HS or LS $\to$ IS $\to$ HS.
In a number of experimental and theoretical papers
\cite{Raccah67,Senaris95,Asai94,Itoh94,Kunes11,Krapek12,Knizek09} their authors proposed
the LS $\to$ HS type scenario to explain the physical properties of RCoO$_3$ cobaltites.
On the other hand, numerous experimental and theoretical studies indicated
that IS state is the closest to the ground LS state, and LS $\to$ IS type scenario
takes place (see e.g. \cite{Zobel02,Baier05,Korotin96,Nekrasov03,Knizek05}).
Thus, the nature and hierarchy of different spin states of RCoO$_3$ remain the subject of
further experimental and theoretical research.

One of the most efficient directions of such research can be a study of high pressure
effect on the magnetic properties of cobaltites.
The first and so far the only investigation of the uniform pressure effect on magnetic
susceptibility of LaCoO$_3$ \cite{Asai97} has revealed a strong decrease in $\chi$ under pressure
and a shift of the characteristic maximum on $\chi(T)$ dependence to higher temperatures.
Later, indirect estimates of pressure effects on susceptibility of LaCoO$_3$ were
obtained from measurements of the volume magnetostriction \cite{Sato08}.
However, the quantitative results of both papers differ significantly and need to be verified.

\begin{figure}[]
\begin{center}
\includegraphics[width=0.45 \textwidth]{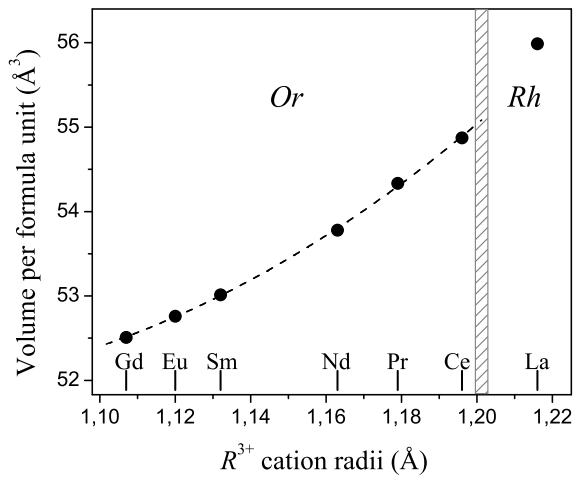}
\caption{Averaged experimental values of the volume per formula unit for RCoO$_3$
compounds at room temperature, according to the data of Refs.
\cite{Baier05,Kobayashi06,Novak13,Ovchinnikov15,Kharko12,Pekinchak14,Kharko14,
Pekinchak15,Liu91,Wang05} as a function of the radius R$^{3+}$ ion \cite{Shannon76}
for the orthorhombic (Or) and rhombohedral (Rh) crystal structure.} \label{V(R)}
\end{center}
\end{figure}

Compared to LaCoO$_3$, a contribution of Co$^{3+}$ ions to the magnetic susceptibility of
the other RCoO$_3$ cobaltites was studied to a lesser extent because of difficulties of its
observation against the background of strong magnetism of rare-earth ions.
Also, to the best of our knowledge, there were no studies of magnetic properties of
these compounds under high-pressure conditions.
Therefore, of particular interest are the studies of dependence of the spin state of cobalt
ions in RCoO$_3$ compounds on changes in the elementary cell volume under the action of a
uniform pressure, as well as a result of the lanthanoid compression effect (Fig. \ref{V(R)})
and thermal expansion.

In this work, we investigated the magnetic susceptibility of RCoO$_3$ cobaltites with
R= La, Pr, Nd, Sm, and Eu in the temperature range $4.2-300$ K, and also under applied
hydrostatic pressure of up to 2 kbar at fixed temperatures $T=78$ and 300 K.
The obtained experimental results are analysed with the use of the two-level model
\cite{Baier05,Zobel02} in terms of a change in population of the excited state of
Co$^{+3}$ ions under the action of temperature and pressure.
The values of the excitation energy $\Delta$ in LaCoO$_3$ and its pressure derivative d$\Delta$/d$P$, resulting from analysis of the experimental data, are supplemented by the
corresponding theoretical estimates obtained by the DFT+U calculations.
Specifically, the fixed spin moment method \cite{Mohn84} was employed
to obtain a volume dependence of the total energy difference between
the Co$^{3+}$ spin states of LaCoO$_3$.

\section{Experimental details and results}

The LaCoO$_3$, PrCoO$_3$, NdCoO$_3$ and SmCoO$_3$ samples were obtained by
solid-state reaction from corresponding rare earth oxides
(La$_2$O$_3$, Pr$_6$O$_{11}$, Nd$_2$O$_3$, Sm$_2$O$_3$) and Co$_3$O$_4$ as initial reagents.
Stoichiometric amounts of initial oxide powders were carefully mixed, pressed
in the pellets and sintered in air at 1473 K for 24 h, followed by re-grinding
and further firing in air at 1473 K for 36 h.
For preparation of EuCoO$_3$, a stoichiometric mixture of Eu$_2$O$_3$ and Co$_3$O$_4$
powders was ball-milled in ethanol for 4 h with 400 rpm, dried and annealed
in air at 1313 K for 18 h.
After cooling the product was repeatedly ball-milled in ethanol for 2 hours,
dried, pressed in the pellet and sintered in air at 1343 K for 40 h.

According to XRD examinations, all as-obtained products showed pure perovskite structure
without detectable amounts of the parasitic phases.
Crystal structure parameters were successfully refined by full-profile Rietveld
technique in rhombohedral space group {\em R-3c} for LaCoO$_3$ and in  orthorhombic
{\em Pbnm} structure for PrCoO$_3$, NdCoO$_3$, SmCoO$_3$ and EuCoO$_3$
\cite{Kharko12,Pekinchak14,Kharko14,Pekinchak15}.

\begin{figure}[]
\begin{center}
\includegraphics[width=0.37 \textwidth]{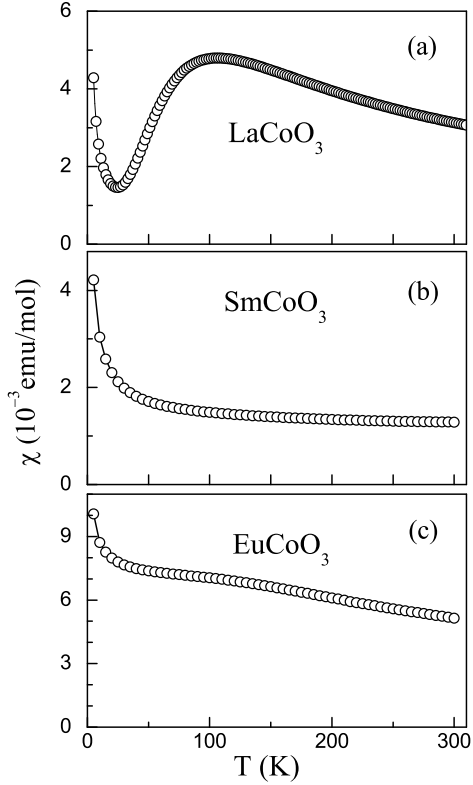}
\caption{Temperature dependences of magnetic susceptibility of RCoO$_3$ compounds
(R= La, Sm, Eu).} \label{X(T)}
\end{center}
\end{figure}

\begin{figure}[]
\begin{center}
\includegraphics[width=0.37 \textwidth]{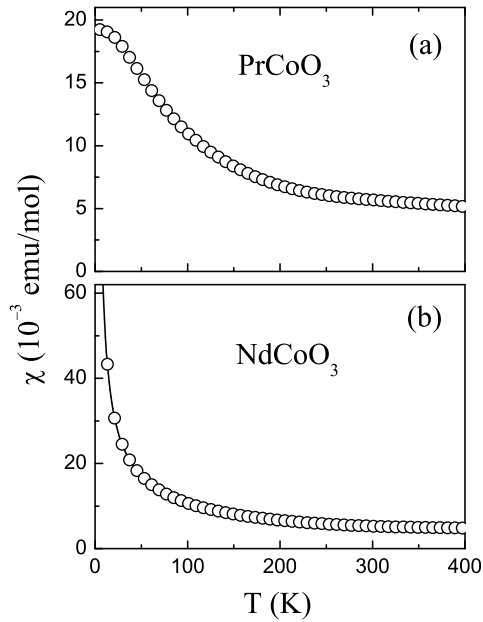}
\caption{Temperature dependences of magnetic susceptibility of
PrCoO$_3$ and NdCoO$_3$.} \label{X(T)2}
\end{center}
\end{figure}

For additional certification of the prepared samples, the temperature dependence of
their magnetic susceptibility was measured in the temperature range 4.2--300 K
in a magnetic field of 1 T using a SQUID magnetometer (for PrCoO$_3$ and NdCoO$_3$,
the temperature interval was 4.2--400 K).
The obtained experimental dependences of $\chi(T)$ are characterized by a noticeable
variety (Figs. \ref{X(T)}, \ref{X(T)2}), and in general are in reasonable agreement
with the literature data for LaCoO$_3$ \cite{Zobel02,Yan04,Baier05},
PrCoO$_3$ \cite{Kobayashi06,Shimokata12,Novak13}, NdCoO$_3$ \cite{Shimokata12,Novak13},
SmCoO$_3$ \cite{Ivanova07} and EuCoO$_3$ \cite{Baier05,Chang03}.
At low temperatures, a noticeable Curie-type contribution due to magnetic impurities was
observed in $\chi(T)$ for RCoO$_3$ compounds with "nonmagnetic" La$^{3+}$ and Eu$^{3+}$ ions.
In the case of R= Sm, Pr, and Nd, the contribution of the magnetic moments of R$^{3+}$ ions
manifests itself significantly in the temperature dependence of susceptibility.

The measurements of the uniform pressure effect on magnetic susceptibility
of RCoO$_3$ compounds (R= La, Nd, Pr, Sm, Eu) were carried out under helium gas
pressure $P$ up to 2 kbar, using a pendulum type magnetometer \cite{Panfilov15}.
The investigated sample was placed inside a small compensating coil
located at the lower end of the pendulum rod.
Under switching on magnetic field $H$, the value of current through the coil,
at which the magnetometer comes back to its initial position,
is the measure of the sample magnetic moment for the fixed value of $H$.
To measure the pressure effects, the pendulum magnetometer was inserted into
a cylindrical non-magnetic pressure cell, which was placed inside a cryostat.
In order to eliminate the effect on susceptibility of the temperature changes
during applying or removing pressure, the measurements were performed at fixed
temperatures 78 and 300 K.
The relative errors of measurements of $\chi$ under pressure did not exceed 0.1\%
for the employed magnetic field $H=1.7$ T.
Within this error, no hysteresis effects were observed in the $\chi(P)$ dependence.

\begin{figure}[]
\begin{center}
\includegraphics[width=0.38 \textwidth]{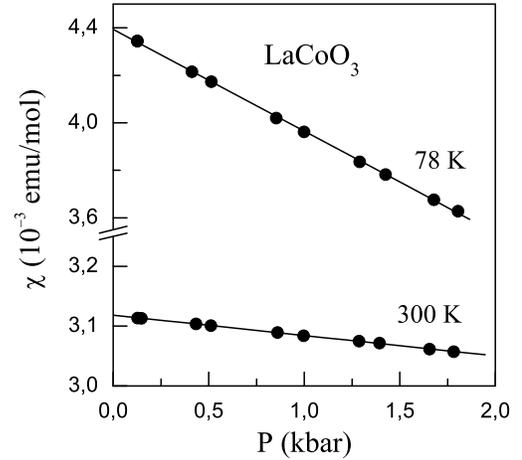}
\caption{Pressure dependence of magnetic susceptibility for LaCoO$_3$ compound at
temperatures 78 and 300 K (the magnitude of the error is less than the size of the symbols).} \label{X(P)1}
\end{center}
\end{figure}

\begin{figure}[]
\begin{center}
\includegraphics[width=0.38 \textwidth]{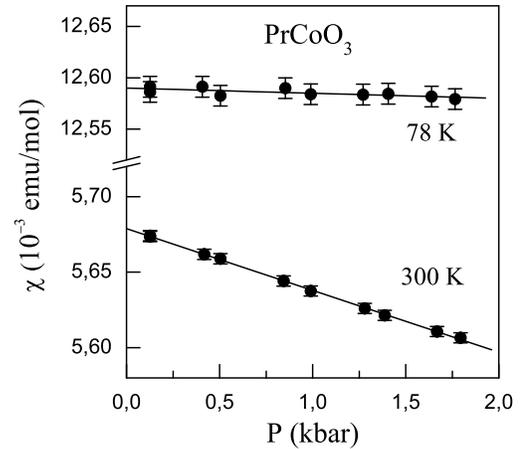}
\caption{Pressure dependence of magnetic susceptibility for PrCoO$_3$ compound at
temperatures 78 and 300 K.} \label{X(P)2}
\end{center}
\end{figure}

\begin{figure}[]
\begin{center}
\includegraphics[width=0.4 \textwidth]{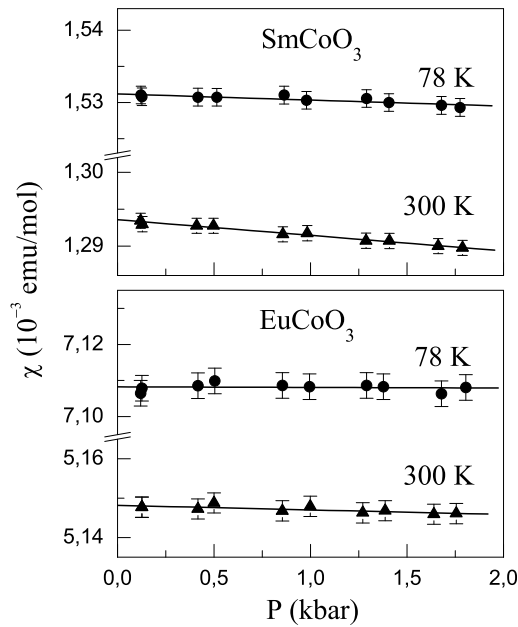}
\caption{Pressure dependence of magnetic susceptibility for SmCoO$_3$ and
EuCoO$_3$ compounds at temperatures 78 and 300 K.} \label{X(P)3}
\end{center}
\end{figure}

The experimental dependences of $\chi(P)$ for the studied cobaltites are presented in
Figs. \ref{X(P)1}, \ref{X(P)2} and \ref{X(P)3}.
As can be seen from Fig. \ref{X(P)1}, a giant decrease in the susceptibility with increasing
pressure is found in LaCoO$_3$ compound, which amounts to about 10\% per kbar at $T=78$ K.
In this case, the pressure effect decreases substantially with increasing temperature,
being about an order of magnitude smaller for $T\simeq 300$ K, but remaining large enough.
On the other hand, in PrCoO$_3$, NdCoO$_3$ and SmCoO$_3$ compounds a noticeable pressure
effect on susceptibility is observed at room temperatures.
For these compounds, the magnitude of pressure effect decreases monotonically along this series.
In the following EuCoO$_3$ compound the pressure effect turns out to be of the same order
as the experimental errors of measurements (Fig. \ref{X(P)3}).
This indicates a weak sensitivity of the dominant Van Vleck paramagnetism in EuCoO$_3$
to changes in the volume of crystal cell.

\begin{table}[h]
\caption{Magnetic susceptibility at $P=0$ (in units of $10^{-3}$ emu/mol) and its
pressure derivative d\,ln$\chi$/d$P$ (Mbar$^{-1}$) for RCoO$_3$ compounds
at temperatures $T=78$ and 300 K}
\vspace{5pt} \label{exp}
\begin{center}
\begin{tabular}{ccccc}
\hline
 Compound &~~~~$\chi$~~~&~d\,ln$\chi$/d$P$~& $~~~~~\chi$~~~~&~d\,ln$\chi$/d$P$~\\
\hline
&\multicolumn{2}{c}{~T=78 K}& \multicolumn{2}{c}{T=300 K}\\
LaCoO$_3$ & 4.40  &~ $ -97 \pm 5$   & 3.11 & $-11.0 \pm 0.5$\\
PrCoO$_3$ & 12.6  &  $~-0.4 \pm0.5$  & 5.68 & $-7.2 \pm 0.5$ \\
NdCoO$_3$ & 12.8  &~ $ -0.7 \pm 0.5$ & 5.35 & $-3.4 \pm 0.5$ \\
SmCoO$_3$ & 1.53  &~ $ -0.5 \pm 0.5$ & 1.29 & $-1.6 \pm 0.5$ \\
EuCoO$_3$ & 7.11  &~ $  0\pm 0.3$    & 5.15 & $-0.2 \pm 0.3$ \\
\hline
\end{tabular}
\end{center}
\end{table}

The obtained pressure derivatives of magnetic susceptibility,
d\,ln$\chi$/d$P\equiv(\Delta \chi/\chi)/\Delta P$ for the studied compounds,
together with the values of $\chi$ at $P=0$ are given in Table \ref{exp}.

\section{Details and results of electronic structure calculations for $\rm LaCoO_3$}

To reveal the origin of the anomalously large pressure effect on magnetic
susceptibility in LaCoO$_3$, we have carried out detailed calculations
of electronic structure for this compound.
As was shown in Refs. \cite{Korzhavyi99,Ravindran02,Harmon13}, the DFT-LSDA
approximation predicts an incorrect metallic ground state of LaCoO$_3$.
For a more adequate description, the DFT+U approach was employed, which has provided
the semiconducting ground state, in agreement with experiments
(see Refs. \cite{Korotin96,Nekrasov03,Knizek05,Spaldin09,Singh17}).

The present calculations of electronic structure for LaCoO$_3$ were performed
using the linearized augmented plane wave method with
a full potential (FP-LAPW, Elk implementation \cite{elk}).
The results of the FP-LAPW method were compared with the calculations
performed by using the Quantum-Espresso code \cite{Giannozzi09,QE}.
We have used the projector-augmented wave (PAW) potentials \cite{Corso14,Topsakal14},
which are directly applicable for the Quantum Espresso code.
The DFT+U approach was employed within the generalized gradient
approximation (GGA) for the exchange-correlation functional \cite{pbe96}.
The effective Coulomb repulsion energy U$_{\rm eff}$=U--J= 2.75 eV was adopted
for Co$^{3+}$ ions according to Refs. \cite{Spaldin09,Singh17},
where such value of U $\equiv$ U$_{\rm eff}$ has provided the correct ground-state
of LaCoO$_3$.

\begin{figure}[]
\begin{center}
\includegraphics[trim=0mm 0mm 0mm 0mm,scale=0.9]{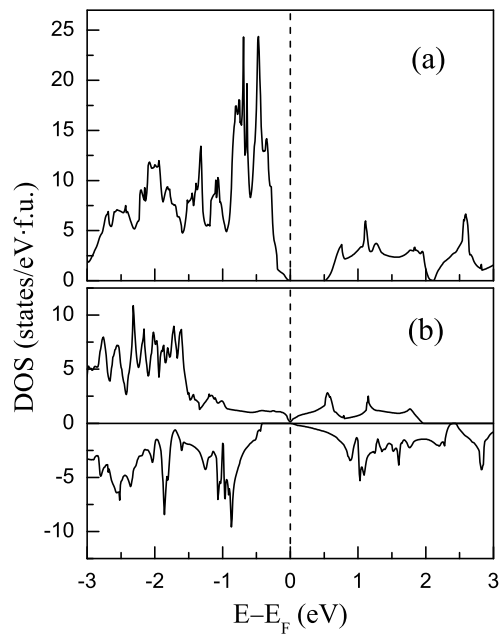}
\caption{(a) -- Density of electronic states for the LS configuration of cobalt ions in LaCoO$_3$; 
(b) DOS for the IS configuration for different spin directions.
The Fermi level is marked by a dashed vertical line.}
\label{DOS}
\end{center}
\end{figure}

The calculated density of electronic states (DOS) for the ground state of LaCoO$_3$
is shown in Fig. \ref{DOS} (a).
According to our DFT+U calculations, the ground state of LaCoO$_3$
is a paramagnetic dielectric with an energy gap of about 0.5 eV,
close to the experimental value of Ref. \cite{Sharma92}.
For this low-spin state of Co$^{3+}$, the valence band is formed by $t_{2g}$ states
of cobalt and $2p$ oxygen orbitals, whereas the conduction band is formed by $e_g$
states of cobalt.
The main features of the calculated electronic structure for the ground state
of LaCoO$_3$ appeared to be in agreement with the results of previous calculations
\cite{Knizek05,Spaldin09,Singh17}.
We have also calculated the volume dependence of the total energy $E(V)$ and obtained
the value of equilibrium volume $V_{\rm th} \cong 56$ \AA$^3$ for the formula unit
of LaCoO$_3$.
This theoretical value of the volume turned out to be slightly larger (about 1.5\%)
than the experimental value at $T=5$ K (see Ref. \cite{Radaelli02}),
presumably due to peculiarities of the applied GGA+U approach.

\begin{figure}[]
\begin{center}
\includegraphics[trim=0mm 0mm 0mm 0mm,scale=0.8]{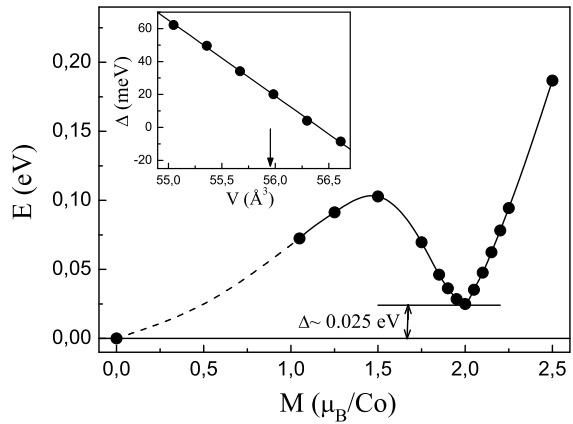}
\caption{Dependence of the total energy on magnetic moment of Co$^{3+}$ ion
calculated by the fixed spin moment (FSM) method
at the theoretical equilibrium volume for LaCoO$_3$.
The energies are given relative to the ground state LS ($M=0$).
In the inset: the volume dependence of the energy difference between the IS and LS states,
$\Delta$; the arrow indicates the value of the theoretical equilibrium volume.}
\label{FSM}
\end{center}
\end{figure}

In order to shed light on the nature of magnetic properties of LaCoO$_3$
we have used the fixed spin moment (FSM) method \cite{Mohn84}.
It was demonstrated (see e.g. Ref. \cite{Mohn06}), that FSM method
gives valuable information on the possible existence and properties of metastable
magnetic phases in solids.
The results of FSM calculations of the total energy $E$ of LaCoO$_3$ as a function
of magnetic moment $M$ for the formula unit are shown in Fig. \ref{FSM}.
As can be seen, there is a clear minimum in the $E(M)$ curve near $M\simeq 2~\mu_{\rm B}$,
which presumably corresponds to the intermediate spin state of the Co$^{3+}$ ion in
LaCoO$_3$ (IS, $S =1)$.
Its energy is only slightly higher ($\simeq 0.025$ eV) than the energy of the ground
low spin state of  Co$^{3+}$ ions (LS, $S=0$), whereas the high-spin state (HS, $S=2$),
according to our estimates, has a much higher energy ($\sim 1$ eV).
Thus, the obtained results indicate that the transition from a non-magnetic to a magnetic
state in LaCoO$_3$ in the region of moderate temperatures actually occurs between
LS and IS state.

We note that for the IS state the spin subbands are shifted relative to one another
and partially overlap, as shown in Fig. \ref{DOS} (b).
Therefore, the calculated ferromagnetic IS state is a half-metal, although
the density of states at the Fermi level is small because of a weak overlap
between the valence band and the conduction band, which actually only touch each other.

We have also calculated the volume dependence of the total energy difference between
the IS and LS states in LaCoO$_3$, $\Delta = E_{\rm IS} -E_{\rm LS}$, for isotropic
volume changes, which is shown in the inset in Fig.~\ref{FSM} and described
by the derivative d$\Delta$/d\,ln$V\simeq-2.5$ eV.
This corresponds to a significant increase in $\Delta$ under pressure,
and, on the other hand, indicates the possibility of the LS--IS spin states crossover
when the volume increases due to thermal expansion.

\section{Discussion}
As already noted, in LaCoO$_3$ compound, where the contribution of cobalt ions
to the magnetic susceptibility is dominant, the evolution of Co$^{3+}$
spin state is the most pronounced,
either as temperature increases, or when pressure is applied.
It is assumed that the unusual temperature dependence of $\chi(T)$ in  LaCoO$_3$
is caused by the temperature-induced transition of Co$^{3+}$ ions from the nonmagnetic
low-spin state LS (S = 0) to a magnetic state with an intermediate spin IS (S=1)
and/or to the high spin state HS (S = 2).
As shown in Ref. \cite{Zobel02}, in the region of low and moderate temperatures,
the dominant contribution to susceptibility $\chi(T)$ of Co$^{3+}$ ions,
$\chi_{\rm Co}(T)$, can be described with the LS$\to$IS transition scenario
by an expression for the two-level system \cite{Baier05,Zobel02,Knizek05}
with the energy difference $\Delta$ for these levels:
\begin{equation}
\chi_{\rm Co}(T) = {{N_{\rm A}g^2\mu^2_{\rm B}S(S+1)}\over{3k_{\rm B}(T-\Theta)}} w(T)\equiv
{C\over (T-\Theta)}w(T).
\label{X(T)_model}
\end{equation}
Here the factor $C/(T-\Theta)$ describes the susceptibility of excited state of
the Curie-Weiss type, $N_ {\rm A}$ is the Avogadro number, $\mu_{\rm B}$ is
the Bohr magneton, $k_{\rm B}$ is the Boltzmann constant, $g$ is the Lande factor,
and $S$ is the spin number.
The factor $w(T)$ determines the temperature dependence of population of the excited state:
\begin{equation}
w(T)={{\nu(2S+1){\rm e}^{-\Delta/T}}\over{1+\nu(2S+1){\rm e}^{-\Delta/T}}},
\label{w(T)}
\end{equation}
where $2S+1$ and $\nu$ are the spin and orbital degeneracy of excited state,
$\Delta$ is the difference between the energies of excited and ground states,
expressed in units of temperature $T$.
We note that in the framework of this approach, a behavior of $\chi_{\rm Co}(T)$
is determined by a single parameter $\Delta$, which depends on temperature.
Starting from the equation (\ref{w(T)}), the $\Delta (T)$ dependence is related
to $w(T)$ as
\begin{equation}
\Delta(T) = T{\rm ln}\left[\nu(2S+1){{1-w(T)}\over w(T)}\right].
\label{Delta(T)}
\end{equation}
In turn, the factor $w(T)$ can be directly determined from the experimental data
using the expression (\ref{X(T)_model}).

Figure~\ref{w,E(T)} (a) shows the evaluated $\chi_{\rm Co}(T)$ dependence for LaCoO$ _3 $.
It was obtained from the experimental $\chi(T)$ dependence (Fig. \ref{X(T)} (a))
by subtraction of the impurity contribution $C/T$ ($C\simeq 20\cdot 10^{-3}$ K$\cdot$emu/mol)
and the temperature independent contribution $\chi_0\simeq 0.35\cdot10^{-3}$ emu/mol,
which describes the total contribution of the diamagnetism of core electrons and
Van Vleck paramagnetism of cobalt ions.
For the analysis of temperature dependence $\chi_{\rm Co}(T)$, we used the usual set
of values for the model parameters \cite{Baier05,Zobel02,Knizek05}: $g=2$, $S=1$,
$\nu=1$ (it is assumed, that the orbital degeneracy of IS state is lifted due to local
distortions of the crystal lattice).
We also expected the paramagnetic Curie temperature $\Theta=0$, assuming that interaction
between the magnetic moments of cobalt ions is negligible.

\begin{figure}[]
\begin{center}
\includegraphics[width=0.4 \textwidth]{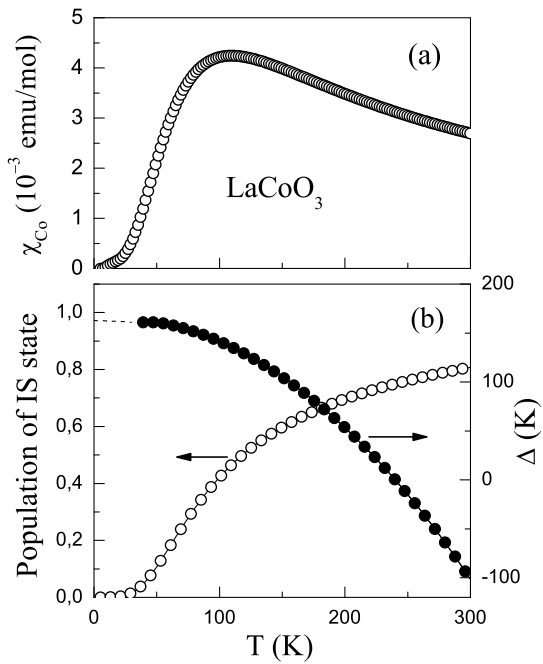}
\caption{Temperature dependences of the contribution of Co$^{3+}$ ions to magnetic
susceptibility of LaCoO$_3$ (a), and the population of excited state, $w(T)$,
and the energy difference $\Delta (T)$ of the IS and LS states (b).}
\label{w,E(T)}
\end{center}
\end{figure}
Using the experimental dependence $\chi_{\rm Co}(T)$ (Fig. \ref{w,E(T)} (a)) and
the expressions (\ref{X(T)_model}), (\ref{Delta(T)}), we have determined
temperature dependences of the population $w(T)$ and the energy difference $\Delta(T)$,
which are shown in Fig. \ref{w,E(T)} (b).
As can be seen, there is a noticeable decrease in $\Delta(T)$ with an increase in
temperature, from $\Delta \simeq 163$ K for $T=0$ K to $\Delta=0$ for $T\simeq 240$ K.
The obtained data on magnitude and temperature dependence of $\Delta$ are close
to the available literature data (see \cite{Knizek05}).

To analyze the experimental data on the effects of pressure, we employed a model description
for the derivative of the magnetic susceptibility with respect to pressure,
d$\chi(T)$/d$P\simeq {\rm d}\chi_{\rm Co}(T)/{\rm d}P$.
According to equations (\ref{X(T)_model}) and (\ref{Delta(T)}) this derivative
can be expressed as follows:
\begin{equation}
{{\rm d}\chi_{\rm Co}(T)\over{\rm d}P} =
-{\chi_{\rm Co}(T)\over T}\left[1-T{\chi_{\rm Co}(T)\over C}\right]{{\rm d}\Delta\over{\rm d}P} .
\label{dX/dP}
\end{equation}
Here the only fitting parameter is the d$\Delta$/d$P$ derivative, and
its choice corresponds to the best agreement of the expression
(\ref{dX/dP}) with experiment.

\begin{figure}[]
\begin{center}
\includegraphics[width=0.4 \textwidth]{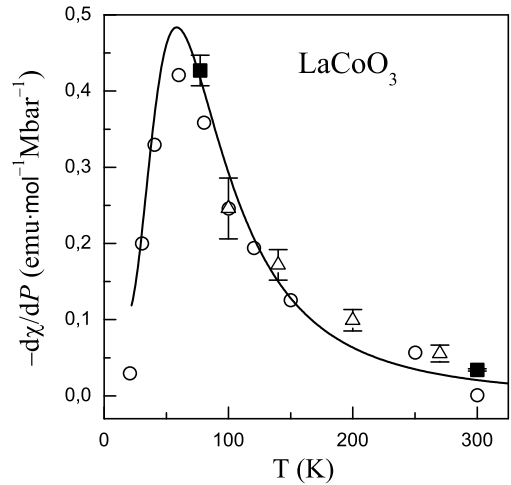}
\caption{Temperature dependence of the derivative of magnetic susceptibility with respect
to pressure, d$\chi$/d$P$, for LaCoO$_3$:
($\blacksquare$) -- experimental data of the present work;
({\large $\circ$}) -- indirect data from the measurements of volume magnetostriction \cite{Sato08};
($\triangle$) -- the chemical pressure effect estimated from the concentration dependence
of the Co$^{3+}$ ion contribution,$\chi_{\rm Co}$, to magnetic susceptibility of  La$_{1-x}$Pr$_x$CoO$_3$ compounds (from Ref. \cite{Kobayashi06});
the solid line is a model description within the expression (\ref{dX/dP}), see text for details.} \label{dchi-dP}
\end{center}
\end{figure}
The obtained in this work experimental values of d$\chi$/d$P$ are presented in
Fig. \ref{dchi-dP}, together with the corresponding indirect estimates
which follow from measurements of the volume magnetostriction $\Delta V/V(H)$ in
LaCoO$_3$ \cite{Sato08}.
These  magnetostriction data are connected with d$\chi$/d$P$ by the relation:
\begin{equation}
{\Delta V\over V}(H)=-{1\over 2V_m}{{\rm d}\chi\over{\rm d}P}H^2,
\label{magnetostriction}
\end{equation}
where $V_m$ and $\chi$ are the molar volume and susceptibility.

The estimates of the chemical pressure effect are also given in Fig. \ref{dchi-dP}.
These estimates were obtained for a number of arbitrarily chosen temperatures
in the range of $100-300$ K, proceeding from the data
of Ref. \cite{Kobayashi06} on the concentration dependence of the Co$^{3+}$ ion contribution,
$\chi_{\rm Co}$, to magnetic susceptibility of isostructural compounds La$_{1-x}$Pr$_x$CoO$_3$
($x=$ 0, 0.1, 0.2, and 0.3) and using relation:
\begin{equation}
{{\rm d}\chi(T)\over{\rm d}P}\simeq{{\rm d}\chi_{\rm Co}(T)\over{\rm d}P}=
-{\partial\chi_{\rm Co}(T)\over\partial x}\left(B{\partial\,{\rm ln}V\over \partial x}\right)^{-1}.
\label{dX/dx}
\end{equation}
Here, the linear decrease in the volume of crystal cell with increasing Pr concentration
amounts to $\partial \,{\rm ln}V/\partial x\simeq-0.03$ \cite{Kobayashi06},
and the bulk modulus $B\simeq 1.35$ Mbar is assumed, which is the mean value ??of
$B\simeq 1.22$ \cite{Zhou05} and 1.5 Mbar \cite{Vogt03}).
As can be seen in Fig. \ref{dchi-dP}, all three sets of experimental data are
in a good agreement with each other, and are reasonably described by using Eq. (\ref{dX/dP})
when choosing d$\Delta$/d$P=+12$ K/kbar.
We note the similarity of the effects of hydrostatic and chemical pressures in
the magnetic susceptibility of LaCoO$_3$, which, apparently, is characteristic
of the entire family of the cobaltites under consideration.

The strong pressure dependence of the parameter $\Delta$, derived from the analysis
of experimental data, also reveals the main mechanism that determines temperature
dependence of $\Delta$ due to the change in volume with thermal expansion.
The corresponding effect in $\Delta (T)$ can be approximately estimated using the expression:
\begin{equation}
\delta\Delta=\Delta(T)-\Delta(0)\approx{\partial\Delta\over\partial\,{\rm ln}V}\times {{V(T)-V(0)}
\over V(0)},
\label{deltaD(T)}
\end{equation}
where $\partial \Delta /\partial \,{\rm ln}V=-B\partial \Delta /\partial P$,
$B$ is the bulk modulus.
Using the values $B\simeq 1.35$ Mbar, $(V(300~{\rm K})-V(0))/V(0) \sim 0.015$
(from Ref. \cite{Radaelli02}) and $\partial \Delta /\partial P\simeq 12\cdot 10^3$
K/Mbar (this work), we have obtained the estimate $\delta \Delta \simeq -240$ K
for $T= 300$ K, in reasonable agreement with the $\Delta (T)$ dependence
in Fig. \ref{w,E(T)} (b).

It should be noted that some improvement in the above used model can be obtained
if one takes into account,
a) the temperature dependence of elastic properties \cite{Naing06},
b) the possible manifestation of interaction between the Co$^{3+}$ moments in
the excited IS state,
and c) the contribution of HS spin state in the high-temperature range,
which were not considered in the preliminary analysis.
Nevertheless, we suppose that these improvements in the model will not lead to
noticeable refinements of the obtained parameters:
\begin{equation}
\Delta\simeq 163~{\rm K}~~\mbox{at}~T= 0~{\rm K, ~~~ d}\Delta/{\rm d}P\simeq 12~{\rm K}/\mbox{kbar}.
\end{equation}

We can also indicate that our estimate of d$\Delta$/d$P$ is close to the value of the chemical
pressure effect, d$\Delta$/d$P\sim 10$ K/kbar, obtained from the analysis of magnetic properties
of La$_{1-x}$Pr$_x$CoO$_3$ system at low concentrations $x$ \cite{Kobayashi06}.

It should be noted that theoretical studies of the volume dependence of $\Delta$
in LaCoO$_3$ also demonstrate a positive pressure effect (see e.g. Refs.
\cite{Korotin96,Kozlenko07}).
However, the available theoretical estimates of the values of both d$\Delta$/d$P$ and
$\Delta$ are significantly different from each other (about an order of magnitude) and
these estimates have to be verified.
On the other hand, the results of our detailed DFT+U calculations of $\Delta$ value
and its volume dependence for LaCoO$_3$ have given the values $\Delta (0)\simeq 250$ K
and d$\Delta$/d$P\simeq 21$ K/kbar, which are in reasonable agreement with the
corresponding estimates obtained in this paper from the analysis of experimental data.
Note that some difference between the calculated and experimental estimates of
d$\Delta$/d$P$ in LaCoO$_3$ may be due to the fact that the calculations were
carried out for volume variations corresponding to a simple change in the scale
of structural parameters, whereas their actual behaviour under uniform pressure
is noticeably anisotropic \cite{Vogt03}.

According to the observed significant increase in excited state energy $\Delta$
in LaCoO$_3$ under pressure, we can expect that in the studied series of RCoO$_3$
compounds (R = Pr, Nd, Sm, Eu) the value of $\Delta$ increases due to a decrease
in the unit cell volume along the R-series (see Fig. \ref{V(R)}).
In particular, for EuCoO$_3$ the decrease in volume per formula unit is about $6\%$
compared to LaCoO$_3$.
With the bulk modulus $B\sim 1.5$ Mbar, this corresponds to a chemical pressure of
about 90 kbar and to increase in $\Delta$ by approximately 1000 K.
Therefore, in RCoO$_3$ compounds we expect a monotonic increase of the characteristic
temperatures, associated with the crossover of spin states,
to ??higher values of $T$.

The temperature dependence of the inverse susceptibility of PrCoO$_3$ and NdCoO$_3$
cobaltites is given in Fig. \ref{Pr,NdCoO3_X(T)-1}.
For the temperature region above 100 K, where the effects of the crystal field are negligible
\cite{Novak13}, this dependence can be described in terms of the Curie-Weiss law:
\begin{equation}
\chi(T)=\chi_0+C/(T-\Theta). \label{CW}
\end{equation}
Here the value $\chi_0 \simeq 0.35\cdot 10^{-3}$ emu/mol is accepted, as in LaCoO$_3 $.
The Curie constants $C$ are close to the values corresponding to the effective magnetic
moments of free Pr$^{3+}$ and Nd$^{3+}$ ions, and $\Theta$ are about $-55$ and $-75$~K
for PrCoO$_3$ and NdCoO$_3$, respectively.
As can be seen in Fig. \ref{Pr,NdCoO3_X(T)-1}, above $T\sim 200$~K for PrCoO$_3$ and
$\sim 240$~K for NdCoO$_3$, there are deviations from the Curie-Weiss law, which can be attributed
to the temperature-induced contribution of the excited states of Co$^{3+}$ ions, $\chi_{\rm Co}$.
At $T=300$~K, the magnitude of these contributions are about 0.72$\cdot 10^{-3}$ and
0.26$\cdot 10^{-3}$ emu/mol for PrCoO$_3$ and NdCoO$_3$, respectively.
Owing to high sensitivity of the $\chi_{\rm Co}$ contribution to pressure,
its existence is also manifested in the experimentally observed increase of
the pressure effect in both cobaltites at room temperature (see Table \ref{exp}).
Estimates of the d$\chi_{\rm Co}$/d$P$ derivative for $T=300$~K can be obtained by
substituting the values of $\chi_{\rm Co}$ in Eq. (\ref{dX/dP}).
They turn out to be equal to $-22.6\cdot10^{-3}$ and $-9.6\cdot10^{-3}$
emu$\cdot$mol$^{-1}$ Mbar$^{-1}$ for PrCoO$_3$ and NdCoO$_3$ when using the value
d$\Delta$/d$P=12\cdot10^3$ K/Mbar, obtained above for LaCoO$_3$ in Eq. (\ref{dX/dP}).
Taking into account these additional contributions to $\chi$, the estimated values
of the total pressure effect are ${\rm d\, ln}\chi/{\rm d}P=-4.4\pm 1.5$ Mbar$^{-1}$
for PrCoO$_3$ and $-2.5\pm 1$ Mbar$^{-1}$ for NdCoO$_3$, which agree well
with the experiment (Table \ref{exp}, $T=300$ K).
Note that the agreement is improved significantly when choosing the value
d$\Delta$/d$P\simeq 18\cdot10^3$ K/Mbar in Eq. (\ref{dX/dP}).

\begin{figure}[]
\begin{center}
\includegraphics[width=0.35 \textwidth]{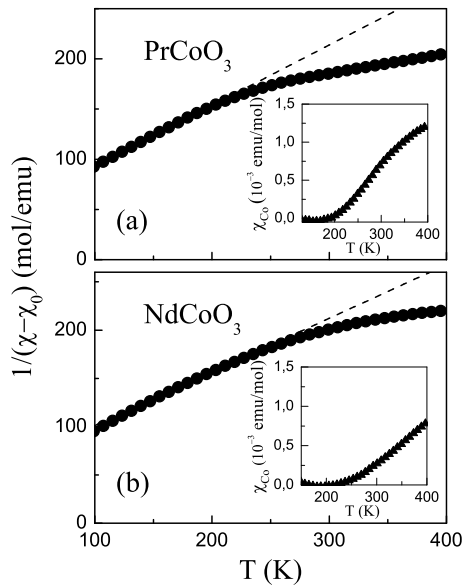}
\caption{Temperature dependence of the inverse susceptibility for PrCoO$_3$ and NdCoO$_3$.
Solid symbols are experimental values, dashed straight lines are approximations by
the Curie-Weiss law.
The inserts show temperature dependence of the contribution to the total susceptibility
of cobalt ions in the excited state (see text for details)} \label{Pr,NdCoO3_X(T)-1}
\end{center}
\end{figure}

Regarding the magnetic properties and pressure effects in the samarium and europium cobaltites,
we note that in RCoO$_3$ series under consideration these compounds have the smallest
volume of the unit cell (Fig. \ref{V(R)}) and, therefore, the highest energy of the excited
Co$^{3+}$ state.
Therefore, in the studied temperature range $4.2-300$~K, a further decrease in the excited magnetism of the Co$^{3+}$ ions is observed, as indicated by a weak growth
of the pressure effect in SmCoO$_3$ at $T=300$ K (Table \ref{exp}) and almost complete
absence of pressure effect in EuCoO$_3$.
The dominant contribution to the magnetism of these compounds is the Van Vleck paramagnetism
of Sm$^{3+}$ and Eu$^{3+}$ ions.
Based on the data obtained in this paper, this type of magnetism is weakly sensitive
to changes in the volume of a crystalline cell under pressure, which is also characteristic
for other paramagnets of this type, e.g. for SmB$_4$ \cite{Panfilov15b}.

\section{Conclusion}
In this work we studied the effects of temperature and hydrostatic pressure on magnetic
susceptibility of RCoO$_3$ cobaltites (where R = La, Pr, Nd, Sm and Eu).
The main aim of this research was to explore dependence of the spin state of cobalt ions
in these compounds on temperature and changes of the unit cell volume both under pressure
and under lanthanide contraction along the RCoO$_3$ series.

The experimental results obtained for LaCoO$_3$, together with the corresponding literature
data on the volume magnetostriction \cite{Sato08} and the chemical pressure effects
\cite{Kobayashi06}, were consistently described in terms of changes in the population
of the excited state of Co$^{3+}$ ions with variations in temperature and pressure, under the assumption that this state corresponds to the intermediate spin of the ions, S = 1.
The analysis using a simple two-level model \cite{Baier05,Zobel02} has revealed an
anomalously large growth under pressure of the relative energy $\Delta$ of the excited state.
The estimated magnitude of this effect, d$\Delta$/d$P\simeq 12$ K/kbar, is reasonably
consistent with the results of our DFT+U calculations of the volume dependence
of $\Delta$ for LaCoO$_3$.
This indicates the adequacy of the DFT+U approach for calculations of the
electronic structure of the cobaltites under consideration.

In our opinion, the obtained behaviour of $\Delta$ under pressure is the main source
of the temperature dependence of this parameter due to the effect of thermal expansion.
It also agrees with the literature data on the continuous decrease in the population of
the excited state of cobalt ions and stabilization of the low-spin state,
which were revealed in the behaviour of the physical properties of LaCoO$_3$
at high pressures (see e.g. Refs. \cite{Vogt03,Vanko06}).
In addition, the existing $\Delta(P)$ dependence explains a shift of the
characteristic temperatures of the spin-crossover transitions in RCoO$_3$
compounds to their higher values due to the chemical pressure under lanthanide contraction.
It should also be noted that the observed similarity between the effects of hydrostatic
and chemical pressures on the value of $\Delta$ and the contribution of cobalt ions to
magnetic susceptibility of cobaltites confirms the strong dependence of the spin state
of Co$^{3+}$ ions on interatomic distance in LaCoO$_3$ and related compounds.

\section{Acknowledgements}
This work was supported by the Russian Foundation for Basic Research
according to the research project 17-32-50015-mol\_nr.
Some authors (L.O.V., V.H., G.E.G. and A.S.P.) acknowledge partial support of 
the Ukrainian Ministry of Education and Sciences under Project ``Feryt''.
The authors acknowledge numerous discussions on the subject with V.V. Efimov.

\end{document}